\documentclass[%
reprint,
superscriptaddress,
 amsmath,amssymb,
 aps, physrev,
]{revtex4-2}
\usepackage{color}
\usepackage{xcolor}
\usepackage{graphicx}% Include figure files
\usepackage{dcolumn}% Align table columns on decimal point
\usepackage{siunitx}
\usepackage{bm}% bold math
\usepackage[version=4]{mhchem}
\usepackage{hyperref}% add hypertext capabilities

\hypersetup{colorlinks=true,linkcolor=blue,citecolor=blue,urlcolor=blue}

\usepackage[colorinlistoftodos]{todonotes}
\usepackage{soul} %\st delete, \hl, highlight, \ul, _, \caps

\begin{document}

\title{Spectroscopic Evidence for Nontrivial Band Topology in Superconducting FeTe/MnTe Heterostructure}

\author{Shiwu Su}
\affiliation{Hefei National Research Center for Physical Sciences at the Microscale, University of Science and Technology of China, Hefei, 230026, China}
\affiliation{School of Emerging Technology, University of Science and Technology of China, Hefei, 230026, China}
\affiliation{Hefei National Laboratory, Hefei 230088, China}
	
\author{Yu Liang}
\affiliation{Hefei National Research Center for Physical Sciences at the Microscale, University of Science and Technology of China, Hefei, 230026, China}

\author{Tongrui Li}
\affiliation{Hefei National Research Center for Physical Sciences at the Microscale, University of Science and Technology of China, Hefei, 230026, China}

\author{Zhen Wang}
\email{wangzhen03@ustc.edu.cn }
\affiliation{Department of Physics, University of Science and Technology of China, Hefei, Anhui 230026, China}
	
\author{Yuzhe Wang}
\affiliation{Hefei National Research Center for Physical Sciences at the Microscale, University of Science and Technology of China, Hefei, 230026, China}

\author{Xianglin Li}
\affiliation{Hefei National Research Center for Physical Sciences at the Microscale, University of Science and Technology of China, Hefei, 230026, China}
\affiliation{School of Emerging Technology, University of Science and Technology of China, Hefei, 230026, China}
\affiliation{Hefei National Laboratory, Hefei 230088, China}

\author{Sen Liao}
\affiliation{Hefei National Research Center for Physical Sciences at the Microscale, University of Science and Technology of China, Hefei, 230026, China}

\author{Pengxu Ran}
\affiliation{Hefei National Research Center for Physical Sciences at the Microscale, University of Science and Technology of China, Hefei, 230026, China}

\author{Jiexiong Sun}
\affiliation{Hefei National Research Center for Physical Sciences at the Microscale, University of Science and Technology of China, Hefei, 230026, China}
\affiliation{School of Emerging Technology, University of Science and Technology of China, Hefei, 230026, China}
\affiliation{Hefei National Laboratory, Hefei 230088, China}

\author{Shengtao Cui}
\affiliation{National Synchrotron Radiation Laboratory, University of Science and Technology of China, Hefei 230026, China}

\author{Zhe Sun}
\affiliation{Hefei National Laboratory, Hefei 230088, China}
\affiliation{National Synchrotron Radiation Laboratory, University of Science and Technology of China, Hefei 230026, China}

\author{Zhengtai Liu}
\affiliation{Shanghai Synchrotron Radiation Facility, Shanghai Advanced Research Institute, Chinese Academy of Sciences, Shanghai 201210, China}
\affiliation{National Key Laboratory of Materials for Integrated Circuits, Shanghai Institute of Microsystem and Information Technology, Chinese Academy of Sciences, Shanghai 200050, China}

\author{Jishan Liu}
\affiliation{Shanghai Synchrotron Radiation Facility, Shanghai Advanced Research Institute, Chinese Academy of Sciences, Shanghai 201210, China}
\affiliation{National Key Laboratory of Materials for Integrated Circuits, Shanghai Institute of Microsystem and Information Technology, Chinese Academy of Sciences, Shanghai 200050, China}

\author{Mao Ye}
\affiliation{Shanghai Synchrotron Radiation Facility, Shanghai Advanced Research Institute, Chinese Academy of Sciences, Shanghai 201210, China}
\affiliation{National Key Laboratory of Materials for Integrated Circuits, Shanghai Institute of Microsystem and Information Technology, Chinese Academy of Sciences, Shanghai 200050, China}

\author{Jing Tao}
\affiliation{Hefei National Research Center for Physical Sciences at the Microscale, University of Science and Technology of China, Hefei, 230026, China}
\affiliation{Department of Physics, University of Science and Technology of China, Hefei, Anhui 230026, China}

\author{Donglai Feng}
\email{dlfeng@hfnl.cn}
\affiliation{Hefei National Laboratory, Hefei 230088, China}
\affiliation{New Cornerstone Science Laboratory, ShanghaiTech University, 200031, Shanghai, China}

\author{Juan Jiang}
\email{jjiangcindy@ustc.edu.cn}
\affiliation{Hefei National Research Center for Physical Sciences at the Microscale, University of Science and Technology of China, Hefei, 230026, China}
\affiliation{School of Emerging Technology, University of Science and Technology of China, Hefei, 230026, China}
\affiliation{Hefei National Laboratory, Hefei 230088, China}

\begin{abstract}

FeTe has long been regarded as a nonsuperconducting antiferromagnetic metal with trivial band topology, but recent advances in stoichiometry control have begun to challenge this picture. Here we use higher-order epitaxy on zinc-blende MnTe to stabilize near-stoichiometric FeTe with strongly suppressed interstitial Fe and a superconducting transition onset near 13~K. Angle-resolved photoemission spectroscopy reveals markedly enhanced quasiparticle coherence, well-defined Fe-derived hole bands, and a nearly two-dimensional Dirac-cone-like state near the Fermi level. First-principles calculations identify an inversion between odd- and even-parity bands, yielding nontrivial $Z_2$ topology and a Dirac surface state consistent with experiment. These results elucidate the intrinsic electronic structure of stoichiometric superconducting FeTe and provide evidence for nontrivial band topology, positioning FeTe/MnTe as a promising platform for exploring topological superconductivity.

\end{abstract}
	
\maketitle
	
%%%%%%%%%%%%%%%%%%%%%%%%%%%%%%%%%%%%%%%%%%%%%%%%%%%%%%%%%%%%%%%%%%%

\textit{Introduction--}The 11-type iron chalcogenide family FeTe$_{1-x}$Se$_x$ provides a structurally simple yet versatile platform for investigating the interplay among electronic correlations, magnetism, superconductivity, and band topology~\cite{ref1,ref2,ref3,ref4,ref5,ref6}. At intermediate compositions, FeTe$_{1-x}$Se$_x$ has emerged as a leading candidate for intrinsic topological superconductivity, with the potential to host Majorana zero modes~\cite{ref3,ref4,ref7}. Toward the FeTe end member, however, the electronic ground state evolves markedly~\cite{ref8,ref9}. FeTe exhibits bicollinear antiferromagnetic (AFM) order and strong orbital-selective electronic correlations~\cite{ref10,ref11}, while also being widely regarded as a prototypical Hund metal~\cite{ref12}. Consistent with its strongly correlated electronic character, previous angle-resolved photoemission spectroscopy (ARPES) measurements revealed broad, incoherent spectral features near the Fermi level ($E_F$) and a pronounced suppression of the $d_{xy}$-derived spectral weight~\cite{ref13,ref14,ref15}. Together with the absence of bulk superconductivity, these observations established the conventional view of FeTe as a strongly correlated AFM metal~\cite{ref16}.

The evolution toward FeTe is also accompanied by a pronounced change in band topology. At intermediate compositions of FeTe$_{1-x}$Se$_x$, the nontrivial topology originates from a band inversion along $\Gamma$--$Z$ between an odd-parity band dominated by chalcogen $p_z$ orbitals and the even-parity $d_{xz}$ band~\cite{ref3,ref4,ref17}. With increasing Te content, enhanced electronic correlations and composition-dependent band shifts drive the $p_z$-dominated band to higher binding energy~\cite{ref3,ref6}. Dynamical mean-field theory calculations therefore suggest that, near the FeTe limit, the $p_z$-dominated band lies entirely below the $d_{xz}$ band along $\Gamma$--$Z$, eliminating the parity inversion and rendering FeTe topologically trivial~\cite{ref6}. Together with the broad and incoherent quasiparticle spectra observed experimentally, this picture has reinforced the conventional view of FeTe as a strongly correlated, topologically trivial AFM metal~\cite{ref18,ref19}.

Recent advances in stoichiometry control have begun to challenge this conventional picture. Interstitial Fe, which is commonly present in FeTe, carries local magnetic moments that couple to the Fe layers and can stabilize bicollinear AFM order. Recent studies have shown that post-growth Te annealing can substantially remove interstitial Fe, suppress AFM order and induce superconductivity in stoichiometric FeTe/SrTiO$_3$ films~\cite{ref20,ref21}. Importantly, the suppression of interstitial Fe also markedly enhances quasiparticle coherence near $E_F$, enabling direct spectroscopic access to the intrinsic low-energy electronic structure of FeTe~\cite{ref21}. These findings raise a central question: does the topologically trivial electronic structure inferred for conventional FeTe remain valid in stoichiometric superconducting FeTe?

In this work, we demonstrate that higher-order epitaxy on zinc-blende MnTe (111) provides an effective route to near-stoichiometric superconducting FeTe (001). Our results suggest that the MnTe template offers a favorable epitaxial environment for stabilizing FeTe close to the 1:1 stoichiometric limit, enabling FeTe/MnTe heterostructures with a superconducting onset temperature ($T_c$) of approximately 13~K. The enhanced quasiparticle coherence enables ARPES to directly resolve the Fe-derived hole bands and an additional two-dimensional Dirac-cone-like state near $E_F$. Density functional theory (DFT) further identifies an inversion between the odd-parity Te $p_z$-derived and even-parity Fe $d_{xz}$-derived bands along $\Gamma$--$Z$, yielding a nontrivial \(Z_2\) invariant (\(Z_2 = 1\)) and a Dirac surface state consistent with experiment. Together, these findings reveal nontrivial band topology in near-stoichiometric superconducting FeTe and motivate further studies of potential topological superconductivity in FeTe/MnTe.

%\begin{figure}[ht]
%\includegraphics[width=8.6cm]{Fig.1_v11.png}
\begin{figure}[t]
\centering
\includegraphics[width=\columnwidth]{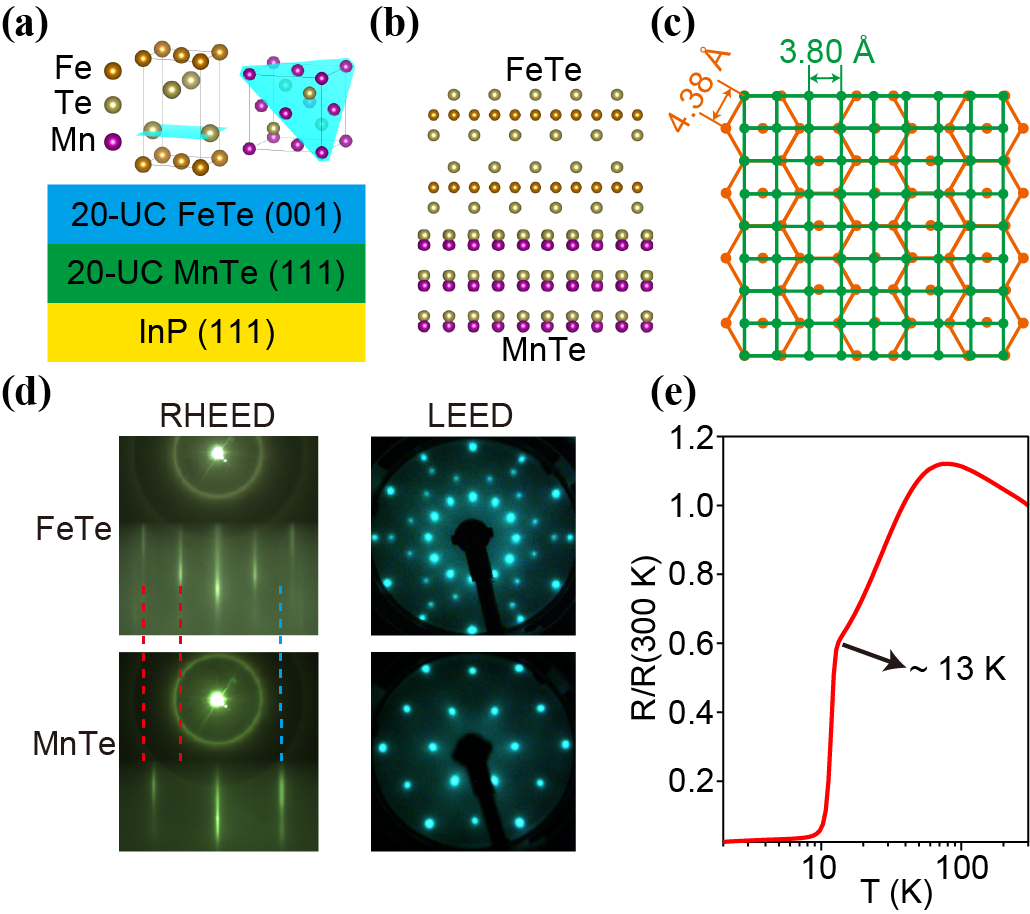}
\caption{\textbf{Epitaxial growth and superconducting transition of the FeTe/MnTe heterostructure.} (a) Schematics of the tetragonal FeTe and zinc-blende MnTe crystal structures (top) and the stacking sequence of the heterostructure (bottom). (b) Cross-sectional schematic of the atomic stacking across the FeTe/MnTe interface. (c) Top-view schematic of the in-plane higher-order epitaxial relationship between FeTe and MnTe. The near-commensurate lattice matching follows \(2a_{\mathrm{FeTe}} \sim \sqrt{3}a_{\mathrm{MnTe}}\). (d) Reflection high-energy electron diffraction (RHEED) patterns of the FeTe and MnTe surfaces (left) and corresponding low-energy electron diffraction (LEED) patterns acquired at 170 eV (right). Red and blue dashed lines mark representative FeTe- and MnTe-related RHEED streaks, respectively. (e) Temperature dependence of the normalized resistance $R/R(300\,\mathrm{K})$, showing a superconducting transition with an onset temperature \ensuremath{\sim}13 K.}

\label{fig1}
\end{figure}

%%%%%%%%%%%%%%%%%%%\section{Results}

\textit{Results--}Figure~\hyperref[fig1]{\ref{fig1}(a)} schematically illustrates the FeTe (001)/MnTe (111) heterostructure fabricated through higher-order epitaxy~\cite{ref22,ref23} (methods in Supplementary Note~1), while Fig.~\hyperref[fig1]{\ref{fig1}(b)} shows the corresponding cross-sectional atomic stacking across the FeTe/MnTe interface. Despite their distinct in-plane symmetries and lattice constants, FeTe and MnTe exhibit a uniaxial higher-order epitaxial relationship enabled by the near-commensurate condition $2a_{\mathrm{FeTe}}\sim\sqrt{3}a_{\mathrm{MnTe}}$~\cite{ref22}, as illustrated in Fig.~\hyperref[fig1]{\ref{fig1}(c)}. A high-quality zinc-blende MnTe (111) film was first grown on InP (111) by substrate-termination-selective epitaxy~\cite{ref24,ref25}, as evidenced by the streaky reflection high-energy electron diffraction (RHEED) pattern and sharp low-energy electron diffraction (LEED) spots shown in Fig.~\hyperref[fig1]{\ref{fig1}(d)}. Upon FeTe (001) deposition, the MnTe-related RHEED streaks gradually disappear, while a new set of sharp streaks characteristic of FeTe emerges. The fourfold-symmetric FeTe domains align along three crystallographically equivalent orientations of the MnTe (111) surface, giving rise to the twelvefold rotational symmetry observed in the LEED pattern. Notably, the resulting FeTe/MnTe heterostructure exhibits a superconducting transition at $T_c\sim13$~K, as revealed by the temperature dependence of the normalized resistance \(R/R(300\,\mathrm{K})\) in Fig.~\hyperref[fig1]{\ref{fig1}(e)} (Supplementary Note~2).

\begin{figure}[t]
\centering
\includegraphics[width=\columnwidth]{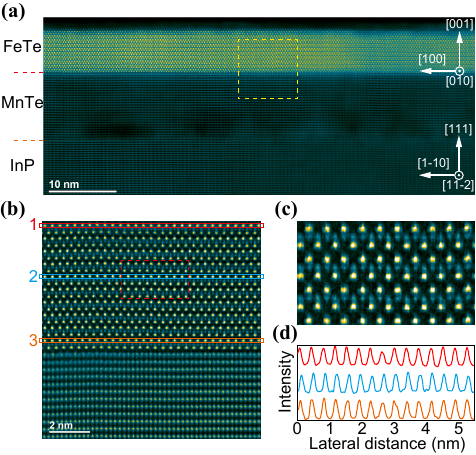}
\caption{\textbf{Cross-sectional atomic structure of the FeTe/MnTe heterostructure.} 
(a) Low-magnification cross-sectional high-angle annular dark-field scanning transmission electron microscopy (HAADF-STEM) image of the FeTe/MnTe heterostructure grown on InP (111). Red and orange dashed lines mark the FeTe/MnTe and MnTe/InP interfaces, respectively. The crystallographic directions of FeTe and MnTe are indicated, with the FeTe and MnTe layers viewed along the [010] and $[11\bar{2}]$ zone axes, respectively. The yellow dashed box indicates the region enlarged in (b). (b) Atomic-resolution HAADF-STEM image across the FeTe/MnTe interface. The red dashed box marks the FeTe region enlarged in (c), while the colored boxes labeled 1--3 indicate the regions used to extract the intensity profiles in (d). (c) Enlarged HAADF-STEM image of the selected FeTe region. Within the experimental resolution, no discernible atomic contrast is observed at the nominal interstitial Fe sites. (d) Line-intensity profiles extracted from regions 1--3 in (b). The profiles are vertically offset for clarity, with colors corresponding to the marked regions. The regularly spaced intensity maxima, without additional peaks attributable to interstitial Fe, support strong suppression of interstitial Fe in the FeTe layer. Scale bars, 10 nm in (a) and 2 nm in (b).}
\label{fig2}
\end{figure}

Given that conventional FeTe has long been regarded as nonsuperconducting, the emergence of superconductivity calls for a closer examination of its microscopic origin and the underlying changes in the FeTe layer. In particular, interstitial Fe has long been recognized as an important factor shaping the ground state of FeTe~\cite{ref49,ref50,ref9}. We therefore examined the structural quality and atomic-scale stoichiometry of the FeTe layer by cross-sectional scanning transmission electron microscopy (STEM). Figure~\hyperref[fig2]{\ref{fig2}(a)} shows a low-magnification cross-sectional STEM image, revealing laterally uniform FeTe and MnTe layers with well-defined interfaces on the InP substrate. At atomic resolution, Fig.~\hyperref[fig2]{\ref{fig2}(b)} further reveals a highly ordered FeTe lattice and an atomically abrupt FeTe/MnTe interface, with the MnTe layer retaining its zinc-blende structure up to the interface. Within the experimental resolution, the enlarged STEM image in Fig.~\hyperref[fig2]{\ref{fig2}(c)} shows no discernible atomic contrast at the expected interstitial Fe sites. Consistently, the line-intensity profiles in Fig.~\hyperref[fig2]{\ref{fig2}(d)} exhibit regularly spaced intensity maxima without additional peaks attributable to interstitial Fe, further supporting the strong suppression of interstitial Fe in the examined regions of the FeTe film.

The strong suppression of interstitial Fe suggests that the FeTe layer in the present heterostructure is close to the 1:1 stoichiometric limit. This observation offers a new perspective on the widespread superconductivity reported in FeTe/telluride heterostructures~\cite{ref22,ref23,ref26,ref27,ref28,ref29,ref30,ref31,ref32}, which has often been attributed to interfacial effects. By comparison, superconductivity in FeTe/SrTiO$_3$ was realized only recently after strong efforts to suppress interstitial Fe~\cite{ref20,ref21}. Together with emerging evidence that stoichiometric FeTe itself can be superconducting, these findings suggest that telluride templates may facilitate superconductivity by helping to stabilize near-stoichiometric FeTe~\cite{ref33}, rather than acting solely through an interfacial mechanism. Telluride-template-assisted stoichiometry control therefore provides an effective route to realizing superconducting FeTe.

\begin{figure*}[t]
\centering
\includegraphics[width=17.8cm]{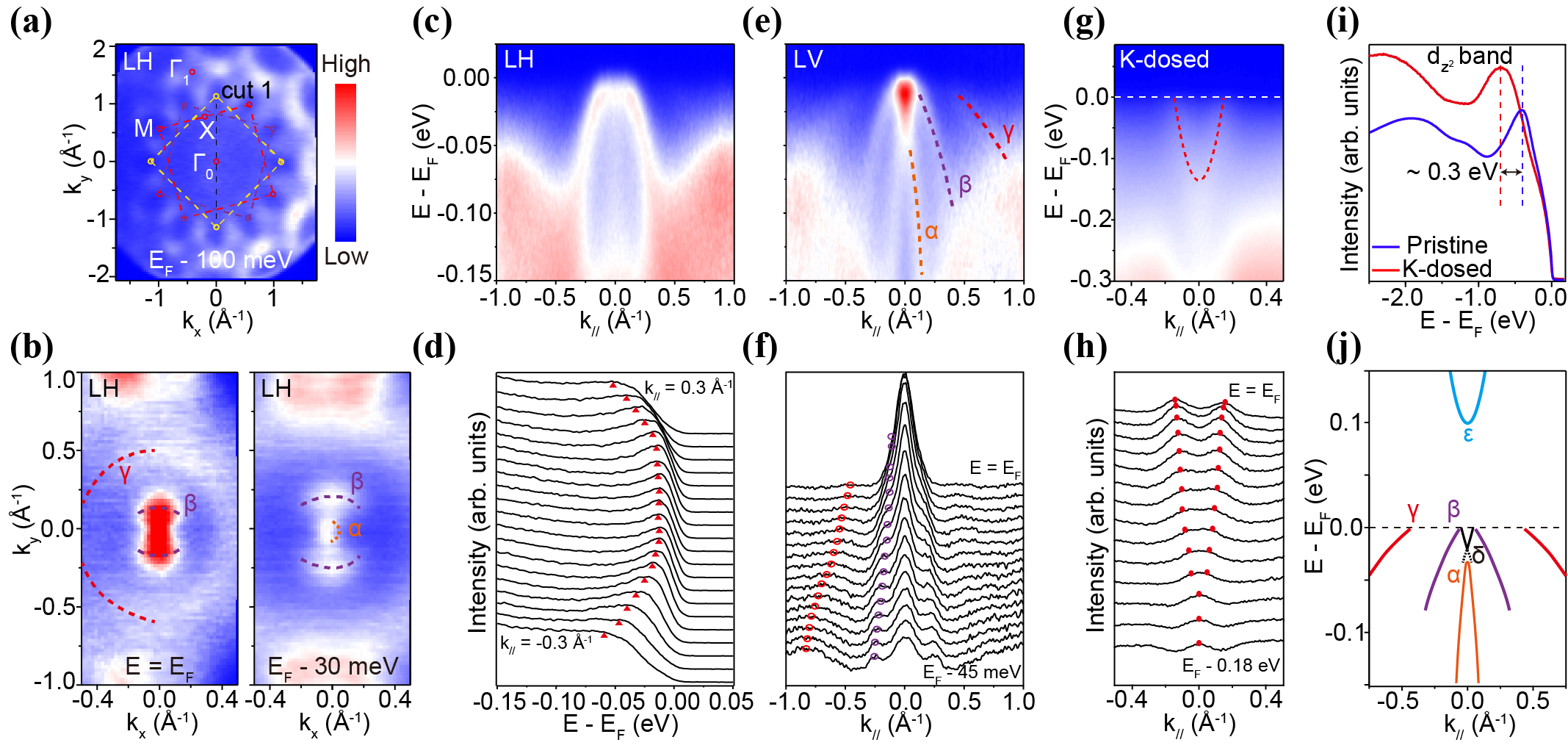}
\caption{\textbf{Low-energy electronic structure of the FeTe/MnTe heterostructure.} 
(a) Constant-energy map acquired with 200 eV LH-polarized photons at a binding energy of 100 meV. Dashed contours and symbols indicate the FeTe Brillouin zones and high-symmetry points. The dashed line labeled “cut 1” marks the momentum cut corresponding to (c) and (d). (b) Constant-energy maps acquired with 64 eV LH-polarized photons and centered at \(E_F\) (left) and a binding energy of 30 meV (right), with the spectral intensity integrated over \ensuremath{\pm}10 meV around each energy. Dashed curves trace the \ensuremath{\alpha}, \ensuremath{\beta} and \ensuremath{\gamma} band contours. (c, d) ARPES spectrum along \ensuremath{\Gamma}--M acquired with 72 eV LH-polarized photons and the corresponding energy-distribution curves (EDCs), with symbols marking the peak positions. (e, f) ARPES spectrum along \ensuremath{\Gamma}--M acquired with 72 eV LV-polarized photons and the corresponding momentum-distribution curves (MDCs). Dashed curves in (e) serve as guides to the eye, and symbols in (f) mark the peak positions. (g) ARPES spectrum near \ensuremath{\Gamma} after K deposition, revealing an electron-like band. The dashed curve serves as a guide to the eye. (h) Corresponding MDCs, with symbols marking the peak positions. The distribution curves in (d), (f), and (h) are vertically offset for clarity. (i) Momentum-integrated EDCs before (blue) and after (red) K deposition, showing an approximately 0.3 eV shift of the \(d_{z^2}\)-derived feature toward higher binding energy. (j) Schematic low-energy band structure of pristine FeTe. The \ensuremath{\delta} and \ensuremath{\varepsilon} bands denote the near-\(E_F\) electron-like feature and the higher-lying conventional electron band, respectively. LH and LV denote linear horizontal and linear vertical polarization, respectively.}
\label{fig3}
\end{figure*}

With interstitial Fe strongly suppressed, we next turn to the low-energy electronic structure of the superconducting FeTe layer. Although FeTe has been extensively studied by ARPES, previous measurements were performed mainly on nonsuperconducting FeTe, whose low-energy spectra are generally broad and incoherent~\cite{ref34,ref35,ref36}. The intrinsic electronic structure of near-stoichiometric superconducting FeTe therefore remains largely unexplored. Figure~\hyperref[fig3]{\ref{fig3}(a)} presents a wide-range constant-energy map acquired with 200~eV photons at a binding energy of 100~meV. The pronounced twelvefold rotational symmetry, consistent with the LEED pattern, arises from the superposition of photoemission signals from three rotational FeTe domains and obscures the electronic structure near the M point. We therefore focus on the electronic states around $\Gamma$, where the nearly isotropic band contours remain distinguishable despite the rotational-domain superposition.

Figure~\hyperref[fig3]{\ref{fig3}(b)} shows constant-energy maps centered at $E_F$ and at a binding energy of 30~meV. The three Fe-derived hole bands are denoted $\alpha$, $\beta$, and $\gamma$ from the innermost to the outermost. One can see that the $\beta$ and $\gamma$ bands cross $E_F$ and form two $\Gamma$-centered hole pockets, whose contours are resolved in the $E_F$-centered map. By contrast, the $\alpha$-band maximum lies below $E_F$, and its contour becomes more clearly visible in the map at 30~meV below $E_F$. Their dispersions are further resolved by energy--momentum cuts along $\Gamma$--$M$ under linear horizontal (LH) and linear vertical (LV) polarizations, as shown in Figs.~\hyperref[fig3]{\ref{fig3}(c)} and \hyperref[fig3]{\ref{fig3}(e)}, respectively. The corresponding energy-distribution curves (EDCs) and momentum-distribution curves (MDCs) in Figs.~\hyperref[fig3]{\ref{fig3}(d)} and \hyperref[fig3]{\ref{fig3}(f)} exhibit well-defined quasiparticle peaks, demonstrating pronounced spectral coherence. Notably, the outermost $\gamma$ band remains well defined and dispersive up to $E_F$. Such coherent low-energy quasiparticle dispersions contrast sharply with the broad and incoherent spectra previously reported for conventional FeTe~\cite{ref11,ref13,ref14,ref15}, providing direct spectroscopic access to the low-energy electronic structure of near-stoichiometric FeTe. The enhanced quasiparticle coherence in the present FeTe/MnTe films is consistent with the strong suppression of interstitial Fe, which in turn substantially suppresses AFM order~\cite{ref20,ref21}.

Beyond these Fe-derived hole bands, we resolve an additional nearly linear band near $E_F$ at $\Gamma$, hereafter denoted the $\delta$ band. To determine whether the $\delta$ band originates from a conventional bulk electron band, we electron-doped the surface via potassium (K) deposition (Supplementary Note~3). As shown in Figs.~\hyperref[fig3]{\ref{fig3}(g)} and \hyperref[fig3]{\ref{fig3}(h)}, a distinct $\Gamma$-centered electron band, hereafter denoted the $\varepsilon$ band, emerges after K deposition, with its dispersion traced by peaks in the corresponding MDCs. Meanwhile, the $d_{z^2}$-derived spectral feature shifts by approximately 0.3~eV toward higher binding energy, as determined from the momentum-integrated EDCs near $\Gamma$ in Fig.~\hyperref[fig3]{\ref{fig3}(i)}, providing an estimate of the doping-induced energy shift. Assuming an approximately rigid-band shift, the minimum of the $\varepsilon$ band in the pristine film is expected to lie well above $E_F$. As summarized schematically in Fig.~\hyperref[fig3]{\ref{fig3}(j)}, the $\varepsilon$ band is therefore energetically distinct from the $\delta$ band observed near $E_F$, arguing against assigning the latter to the conventional electron band.

\begin{figure*}[t]
\centering
\includegraphics[width=17.8cm]{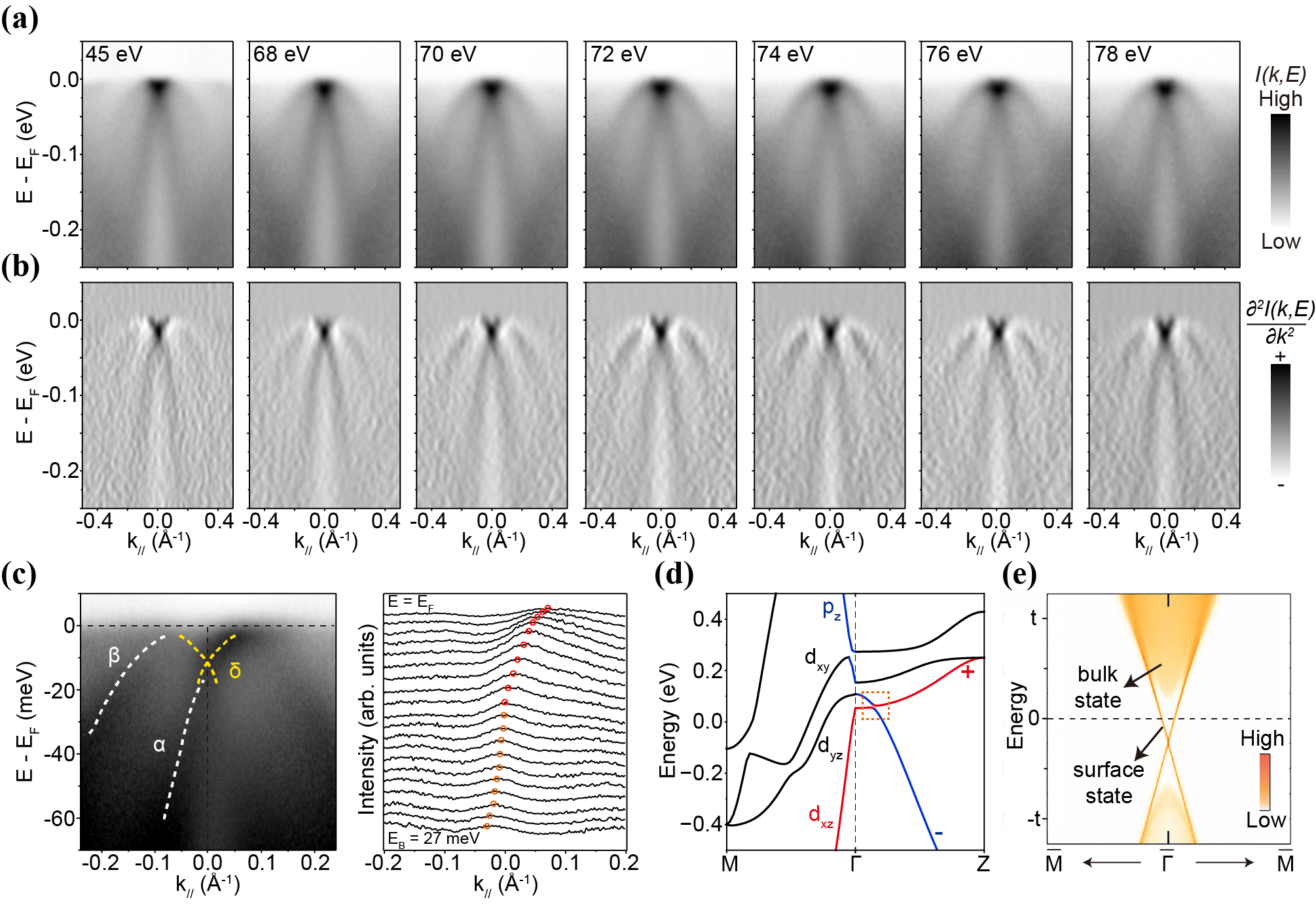}
\caption{\textbf{Two-dimensional Dirac-like state and calculated nontrivial band topology in superconducting FeTe.} 
(a) Synchrotron ARPES spectra along \ensuremath{\Gamma}--M acquired at photon energies from 45 to 78 eV. (b) Corresponding second-derivative plots, highlighting the nearly photon-energy-independent Dirac-cone-like dispersion near \(E_F\). (c) High-resolution laser-ARPES spectrum near \ensuremath{\Gamma} (left) and corresponding MDCs (right). Dashed curves guide the dispersions, and circles mark the MDC peak positions. The MDCs are vertically offset for clarity. (d) Calculated orbital-projected bulk band structure along M--\ensuremath{\Gamma}--Z. The dashed box highlights the band inversion between the even-parity Fe \(d_{xz}\)-derived and odd-parity Te \(p_z\)-derived bands along \ensuremath{\Gamma}--Z. The + and \ensuremath{-} signs denote their parity eigenvalues. Parity analysis yields a nontrivial bulk invariant, \(Z_2\) = 1. (e) Calculated surface spectral function along \(\bar{M}\)--\ensuremath{\bar{\Gamma}}--\(\bar{M}\), showing the projected bulk states and a Dirac-like surface state centered at \ensuremath{\bar{\Gamma}}.}
\label{fig4}
\end{figure*}

Having distinguished the $\delta$ band from the conventional $\varepsilon$ electron band, we next investigate its dimensionality and possible surface-state character. The location and dispersion of the $\delta$ band closely resemble those of the Dirac surface state previously reported in FeTe$_{1-x}$Se$_x$~\cite{ref37}. We therefore performed photon-energy-dependent ARPES measurements to examine its $k_z$ dependence. As shown in Fig.~\hyperref[fig4]{\ref{fig4}(a)}, the $\delta$ band persists over a broad photon-energy range from 45 to 78~eV, despite variations in spectral intensity (Supplementary Note~4). The corresponding second-derivative spectra in Fig.~\hyperref[fig4]{\ref{fig4}(b)} further highlight its dispersion. No discernible $k_z$ dispersion is observed, indicating a predominantly two-dimensional character. High-resolution laser-ARPES in Fig.~\hyperref[fig4]{\ref{fig4}(c)} further resolves the Dirac-cone-like dispersion centered at $\Gamma$, with the corresponding MDC peaks tracing its dispersion. Together, these observations support the identification of the $\delta$ band as a two-dimensional Dirac-like state and motivate further investigation of its possible topological origin.

\begin{table*}[t]
\centering
\caption{Parity analysis of the occupied Kramers pairs at the eight
time-reversal-invariant momenta (TRIMs). The numbers denote odd/even
parity pairs for the occupied states up to band index 24, corresponding
to the $\alpha$ band.}
\label{tab:parity24}

\renewcommand{\arraystretch}{1.15}
\setlength{\tabcolsep}{3pt}
\setlength{\doublerulesep}{2pt}

\begin{tabular*}{\textwidth}{
    @{\extracolsep{\fill}} l *{8}{c} c @{}
}
\hline\hline
\multicolumn{10}{c}{
    Calculated parity at eight TRIMs
} \\[3pt]
\hline

& \multicolumn{8}{c}{
    Number of Kramers-degenerate pairs of odd parity/even parity
} & \\[3pt]
\cline{2-9}

Band index
& $(0,0,0)$
& $(0,0,\pi)$
& $(0,\pi,0)$
& $(0,\pi,\pi)$
& $(\pi,0,0)$
& $(\pi,0,\pi)$
& $(\pi,\pi,0)$
& $(\pi,\pi,\pi)$
& $Z_2$ \\[3pt]
\hline

24 ($\alpha$)
& $5/7$
& $6/6$
& $6/6$
& $6/6$
& $6/6$
& $6/6$
& $6/6$
& $6/6$
& $1$ \\[3pt]
\hline\hline
\end{tabular*}

\end{table*}

To examine the possible topological origin of the $\delta$ band, we performed DFT calculations of the bulk and surface electronic structures (Supplementary Note~5). Previous calculations have suggested that increasing Te content enhances the interlayer overlap of chalcogen $p_z$ orbitals, imparting strong $k_z$ dispersion to the odd-parity $p_z$-derived band and shifting it toward lower energy~\cite{ref3,ref19}. This evolution was proposed to eliminate the band inversion near the FeTe limit, thereby driving the system into a topologically trivial phase. In contrast, our calculated bulk band structure in Fig.~\hyperref[fig4]{\ref{fig4}(d)} shows that, even at the FeTe limit, the $p_z$-derived band remains above the $d_{xz}$-derived band at $\Gamma$ and crosses it along $\Gamma$--$Z$, resulting in a reversal of their relative ordering between $\Gamma$ and $Z$. To determine the bulk topological character, we applied the Fu--Kane parity criterion to the occupied Kramers pairs up to band index 24, corresponding to the $\alpha$ band, at all eight time-reversal-invariant momenta~\cite{ref38}. As summarized in Table~I, the parity analysis yields $Z_2 = 1$, indicating a nontrivial bulk topology. This result contrasts with the trivial band ordering previously proposed for the FeTe limit. Consistent with the nontrivial bulk topology, the calculated surface spectral function in Fig.~\hyperref[fig4]{\ref{fig4}(e)} reveals a Dirac surface state at the surface Brillouin-zone center, in agreement with the two-dimensional Dirac-cone-like $\delta$ band observed near $E_F$ by ARPES.

Taken together, our results suggest that telluride templates may play a broader role in FeTe than previously appreciated. Rather than inducing superconductivity solely through interfacial electronic effects, they may also facilitate the stabilization of FeTe close to the stoichiometric limit. In this picture, superconductivity previously reported in a variety of FeTe/telluride heterostructures may, at least in part, originate from the FeTe layer itself. The present FeTe/MnTe system further provides evidence that near-stoichiometric superconducting FeTe can host nontrivial band topology, thereby linking stoichiometry control, superconductivity, and topology within a common framework. Future high-resolution ARPES measurements of the superconducting gap on the candidate topological surface state, together with scanning tunnelling microscopy and spectroscopy searches for zero-energy vortex-bound states, will be important for determining whether the nontrivial surface state participates directly in superconductivity and for assessing the possible emergence of topological superconductivity in FeTe.

\textit{Conclusion--}To summarize, by combining telluride-interface engineering with Te-rich molecular beam epitaxy, we realize superconducting FeTe with sufficient quasiparticle coherence for direct ARPES investigation. ARPES resolves three Fe-derived hole bands and a two-dimensional Dirac-cone-like state near $E_F$, while DFT reveals a band inversion between odd- and even-parity states along $\Gamma$--$Z$, yielding a nontrivial $Z_2$ invariant. The convergence of experiment and theory revises the conventional picture of FeTe as a nonsuperconducting, topologically trivial endpoint and positions stoichiometric FeTe as a promising platform for exploring topological superconductivity.

\textit{Note added:} Recently, we became aware of an independent ARPES study that also reported a Dirac-cone-like surface state in stoichiometric FeTe films grown on SrTiO$_3$ substrates~\cite{ref48}.

\textit{Acknowledgments--}This work is supported by the National Natural Science Foundation of China (Grant No. 12574071), the National Key R\&D Program of China (Grant No. 2023YFA1406304), Quantum Science and Technology-National Science and Technology Major Project (No. 2021ZD0302803), and the New Cornerstone Science Foundation. Part of this research used Beamline 03U of the Shanghai Synchrotron Radiation Facility, which is supported by the ME2 project under contract no. 11227902 from National Natural Science Foundation of China.

\bibliographystyle{apsrev4-2}  
\bibliography{references}
	%\begin{thebibliography}{10}
		%\begin{references}

	%\end{thebibliography}
\end{document}